\documentstyle[12pt]{article}
\newcommand\be{\begin{equation}}
\newcommand\bear{\begin{eqnarray}}
\newcommand\ee{\end{equation}}
\newcommand\eear{\end{eqnarray}}

\def\pa {{\partial}}
\def\al {{\alpha}}

\def\kap {{\kappa}}
\def\lm {{\lambda}}
\def\Om {{\Omega}}
\def\yp {{\varepsilon}}
\def\ypd {{\dot\varepsilon}}
\def\yps {{\varepsilon^*}}

\def\alb {{\bf \alpha}}
\def\beb {{\bf \beta}}
\def\gmb {{\bf \gamma}}
\def\deb {{\bf \delta}}
\def\Ac {{\bf \cal A}}
\def\Ab {{\bf \alpha}}
\def\Bb {{\bf B}}
\def\Bc {{\bf \cal B}}
\def\Xb {{\bf X}}
\def\Pb {{\bf P}}
\def\Pc {{\bf \cal P}}
\def\Qb {{\bf Q}}
\def\fb {{\bf f}}
\def\bb {{\bf b}}
\def\ab {{\bf a}}
\def\mb {{\bf m}}
\def\nb {{\bf n}}
\def\pb {{\bf p}}
\def\qb {{\bf q}}
\def\ub {{\bf u}}
\def\xb {{\bf x}}
\def\yb {{\bf y}}
\def\zb {{\bf z}}
\def\Cb {{\bf C}}
\def\1b {{\bf 1}}
\def\0b {{\bf 0}}
\def\Scs {{Schr\"{o}dinger cat states}}

 INFN-NA-IV-93/49~~~~~~~~~~~~~~~~~~~~~~~~~~~~~~~~~~~~~~~DSF-T-93/49

\title{\bf
Even and odd coherent states\\ (Schr\"{o}dinger cat states)\\ for
multimode parametric systems}
 \author{ V.~V.~Dodonov \\
{\em Moscow Institute of Physics and Technology, Zhukovsky, 140160
Russia} \\[3mm]
V.~I.~Man'ko \\
{\em Lebedev Physics Institute, Moscow,
Russia} \\ and  \\
{\em Departamento di Scienze Fisiche,} \\ {\em Universita di Napoli
"Federico II"} \\ and \\ {\em INFN, Sezione di Napoli} \\
{\em Mostra d'Oltremare, Pad. 20-80125
Napoli, Italy} \\[3mm]
D.~E.~Nikonov \\
{\em Department of Physics, Texas A{\rm \&}M University,} \\
{\em College Station, TX 77843-4242 } \\ and \\
{\em Moscow Institute of Physics and Technology,} \\ {\em Zhukovsky, 140160
Russia}   }

\begin{document}

\baselineskip=24pt

\maketitle

\begin{abstract}
\baselineskip=24pt

The multimode even and odd coherent states (multimode
Schr\"{o}dinger cat states) are constructed for polymode
parametric oscillators of the electromagnetic field.
The evolution of the photon distribution
function is evaluated explicitly. The
distribution function is expressed in terms of the
multivariable Hermite polynomials, its means and dispersions are
calculated. The conditions for the existence
of squeezing are formulated.
The correlations among the different modes of \Scs \ are studied.
\end{abstract}
\vskip 0.2 cm
\hskip 1cm  PACS Numbers 03.65.-w, 42.50.+q
\vskip 0.5 cm

\newpage

\section{Introduction}

Even and odd coherent states for one-mode systems have been
introduced in \cite{Dod72}. They have been shown to represent a special
set of nonclassical states by Nieto and Truax in Ref.~\cite{Nie93}.
These states have been considered in Ref.~\cite{Per}.
Their properties have been studied in
\cite{BuzekVid}-\cite{Hillery}. The even coherent states are similar to
squeezed vacuum states \cite{Holl79}, since they are the superposition
of the photon number states with the even numbers of quanta. Due to this,
the even coherent light may be used
in
interferometric gravitational wave detectors to give the same
effect of increasing the sensitivity of these devices, which could be
produced by the replacement of the vacuum state by the squeezed
vacuum light at the unused port of the interferometer \cite{Sol93}.
The photon statistics of one-mode even and odd coherent states has the
properties of that of nonclassical light \cite{BuzekVid}.

The even and odd coherent states (\Scs\ \cite{Schr35},\cite{Yur86})
may be generated
in different processes \cite{Gea}-\cite{Gerry93}.
Gea-Banacloche \cite{Gea} showed a
possibility of appearing \Scs\ in resonant Jaynes-Cummings model. Gerry
and Hach \cite{Gerry93}
demonstrated a possibility to generate even and odd coherent
states  for long-time evolution of the competition between a two-photon
absorbtion and two-photon parametric processes for a special initial
field state. Recently a possibility to generate quantum
superposition of macroscopically distinguishable states in two
cavities was shown \cite{Dav93}.
In Ref.~\cite{Akul} methods of engineering
different kinds of field states  (including \Scs ) were suggested.

Multimode even and odd
coherent states and the photon statistics for these states
have been considered recently \cite{Ansari}. These states have been
constructed as the superpositions of the usual multimode coherent
states (which have no squeezing in each coherent component of the
superposition). These states describe the quantum superposition of
macroscopically distinguishable states for systems with many degrees
of freedom. The nonclassical states of light corresponding to multimode
field are interesting since they may be generated by similar methods
discussed for single-mode case.

Agarwal, Graf, Orszak, Scully and Walther
have studied \cite{Graf} the possibility of generating
two-mode cat states by a continuous measurement.

 The photon distribution function for the multimode
\Scs\  turned out to be a deformed product of Poisson distribution
functions describing the contribution of each mode. But this function
is not factorised into the product of one-dimensional Poisson distributions.
This fact
demonstrates statistical correlations existing among the modes
of the multimode {\Scs} \cite{Ansari}.
In this article we address the following problem.
How does
the squeezing in the multimode coherent components of the
\Scs\  influence these correlations among
the different modes?

So the aim of this work is to study the photon statistics of
multimode even and odd light taking into account
the squeezing in the multimode coherent components through which the \Scs\
are
expressed. We construct the multimode photon distribution function and
calculate the means and dispersions related to this distribution.
Moreover, we consider the evolution of one-mode
and multimode even and odd coherent states due to parametric excitation
of a quantum system, since this process is known to produce
the squeezing in the quadratures of initially coherent light (see, for example,
\cite{183}).
It may correspond to the
evolution of the \Scs \  in a resonator filled with a medium which
parameters vary in time
\cite{Media}.
We shall demonstrate that
the multimode \Scs \ of Ref.~\cite{Ansari} turn into a multimode
squeezed {\Scs}, since
each coherent component determining the even and odd coherent states
transforms into a multimode squeezed and correlated state.

The material of the article is organised in the following manner.
In Section~2 we review the properties of the multimode \Scs \  discussed
in Ref.~\cite{Ansari}. In Section~3 the properties of polymode parametric
quadratic systems are considered. As an example of
a parametric one-mode system,
the \Scs \ of an electromagnetic field oscillator are studied in Section~4.
The multimode squeezed \Scs\  resulting from the time evolution of the \Scs
\  of
a parametric electromagnetic field oscillator are presented in Section~5.
We generalize the above consideration to the case of
external driving force in Section~6.

\section{Multimode even and odd coherent states}

For each mode of the electromagnetic
field, the eigenvectors of its annihilation
operator $a_j$ (coherent states) with the eigenvalue
$\alpha_j$ are generated from the vacuum state $|0>$
by the displacement operator \cite{Gla63}, i.e.,
\be
|\alpha_j\rangle = D(\alpha_j) |0\rangle \equiv
\exp (\alpha_j a_j^{\dag} - \alpha_j^* a_j) |0\rangle.
\ee
In \cite{Ansari} N-mode even and odd
 coherent states defined as
\be
|\Ab_\pm\rangle = N_\pm ( |\Ab\rangle \pm |-\Ab\rangle )
\label{cat}
\ee
were considered, where $|\Ab\rangle$ is a direct product
of coherent states in each mode:
\be
|\Ab\rangle = |\al_1,\al_2,\dots,\al_N \rangle .
\ee
The normalization constants
\be
N_+ = \frac { \exp(|\Ab|^2/2) } { 2\sqrt{\cosh|\Ab|^2} }
\ee
\be
N_- = \frac { \exp(|\Ab|^2/2) } { 2\sqrt{\sinh|\Ab|^2} }
\ee
contain the square of the parameter complex vector
$\Ab = (\al_1,\al_2,\dots,\al_N)$. The presence of this
non-factorizable
factor expresses statistic correlations between the modes.

The action rules for annihilation operators on these
states
\be
a_j|\Ab_+\rangle = \al_j\sqrt{\tanh|\Ab|^2} |\Ab_-\rangle ,
\label{rul+}
\ee
\be
a_j|\Ab_-\rangle = \al_j\sqrt{\coth|\Ab|^2} |\Ab_+\rangle ,
\label{rul-}
\ee
allow us to calculate means and dispersions of quadratures and photon
numbers.
The states $|\Ab_+\rangle$ and $|\Ab_-\rangle$ are orthonormal.
Thus using Eq.~(\ref{rul+}) and (\ref{rul-}) we get
\be
\langle\Ab_\pm| a_j |\Ab_\pm\rangle =
\langle\Ab_\pm| a_j^{\dag} |\Ab_\pm\rangle = 0.
\ee
So the multimode dispersions coincide with the second moments
for the operators
\be
\langle\Ab_\pm| a_ja_k |\Ab_\pm\rangle = \al_j\al_k =
\langle\Ab_\pm| a_j^{\dag} a_k^{\dag} |\Ab_\pm\rangle^*.
\ee
Further, the expressions
for other elements of the dispersion matrix are
\be
\sigma_+(a_j^{\dag} a_k) \equiv
\langle\Ab_+| \frac{1}{2}(a_j^{\dag} a_k + a_ka_j^{\dag}) |\Ab_+\rangle
 = \al_j^*\al_k \tanh|\Ab|^2 + \frac{1}{2}\delta_{jk},
\ee
\be
\sigma_-(a_j^{\dag} a_k) \equiv
\langle\Ab_-| \frac{1}{2}(a_j^{\dag} a_k + a_ka_j^{\dag}) |\Ab_-\rangle
 = \al_j^*\al_k \coth|\Ab|^2 + \frac{1}{2}\delta_{jk}.
\ee
The mean photon numbers $n_j = a_j^{\dag} a_j$ prove to be
\be
\langle\Ab_+| n_j |\Ab_+\rangle = |\al_j|^2 \tanh|\Ab|^2,
\ee
\be
\langle\Ab_-| n_j |\Ab_-\rangle = |\al_j|^2 \coth|\Ab|^2.
\ee
Dispersions for the photon numbers
\be
\sigma_{jk\pm} = \langle\Ab_\pm| n_jn_k |\Ab_\pm\rangle
- \langle\Ab_\pm|n_j|\Ab_\pm\rangle \langle\Ab_\pm|
n_k|\Ab_\pm\rangle
\ee
can be expressed as follows
\be
\sigma_{jk+} = |\al_j|^2 |\al_k|^2 {\rm sech}^2|\Ab|^2
+ |\al_j|^2 \tanh|\Ab|^2 \delta_{jk} ,
\ee
\be
\sigma_{jk-} = -|\al_j|^2 |\al_k|^2 {\rm cosech}^2|\Ab|^2
+ |\al_j|^2 \coth|\Ab|^2 \delta_{jk} .
\ee
Comparing the mean value of the square of the photon number
$\sigma_{jj}$ at some fixed $j$ with the mean value of
the photon number in the same mode $n_j$, we see that
for the even cat state the photon statistics is always
superpoissonian and for the odd cat states it is always
subpoissonian. For one-mode \Scs \ this result has been discussed
in Ref.~\cite{BuzekVid}.

\section{Quadratic parametric systems}
\label{se:pasy}

Let us consider systems with generic quadratic hermitian Hamiltonian
\cite{JMP73}, \cite{183.3}
\be
H \ = \ \frac {1}{2} \qb\Bb (t)\qb  .
\ee
Here $\qb$ is a Schr\"{o}dinger vector operator
$(\pb,\xb) = (p_1,p_2,\dots,p_N,x_1,x_2,\dots,x_N)$
composed of the momentum and coordinate operators for each mode.

Coefficients $B_{\mu\nu} = B_{\nu\mu}$ of the
$2N\times 2N$-matrix $\Bb (t)=\Vert B_{\mu\nu}\Vert$
may be arbitrary functions of time.
In this section we will give a review of the results of
Ref.~\cite{183.3} for this system.

Canonical coordinates are expressed through the annihilation and creation
operators as follows
\be
\left( \begin{array}{c}
\pb \\ \xb
\end{array} \right) =
\ub
\left( \begin{array}{c}
\ab \\ {\ab^{\dag}}
\end{array} \right) ,
\label{rozhd}
\ee
where $N \times N$ block-matrices
\be
\ub_{11} = -\ub_{12} = -i\ub_{21} = -i\ub_{22} = -i\1b .
\ee

The evolution operator $U(t)$ transforms the state vector
at zero time to that at time $t$
\be
|\psi (t)\rangle = U(t) |\psi(0)\rangle .
\ee
We define the integral of motion in the
Schr\"{o}dinger picture
as an operator which may  depend on time but
mean values of
which do not vary in the process of time evolution \cite{183}.
Such integrals of motion which describe
the initial momenta and coordinates $\Qb(t) = (\Pb(t),\Xb(t))$ are given by
the relation
\be
\Qb(t) = U(t)\qb U^{-1}(t).
\label{evol}
\ee
Then if it is possible to find the explicit expression
for this operator
\be
\Qb(t) = \Lambda(t)\qb
\ee
from the equation which it obeys
\be
i\hbar \frac{\pa\Qb}{\pa t} = [H,\Qb ] ,
\ee
one can solve Eq.~\ref{evol}
for the evolution operator.
Introducing the $2N\times 2N$
commutator matrix $\Sigma$ by the expression
\be
[q_\mu,q_\nu] = -i\hbar (\Sigma)_{\mu\nu}
\ee
\be
\Sigma = \left(
\begin{array}{cc}
\0b & \1b \\ -\1b & \0b
\end{array}
\right) ,
\ee
where ${\bf 1}$ is an $N \times N$ unity matrix,
we obtain equations for the coefficient
matrix
\be
\dot\Lambda = \Lambda\Sigma B
\ee
with the initial condition $\Lambda (0)$ being the unity matrix.
Provided  this solution is expressed in terms of $N \times N$ block-
matrices
\be
\Lambda =
\left(
\begin{array}{cc}
\lm_1 & \lm_2 \\
\lm_3 & \lm_4
\end{array},
\right)
\label{lambda}
\ee
one can write down the evolution operator in coordinate
representation (Feynman propagator) according to
the formula (2.47) of Ref.~\cite{183.3}
as follows
\be
G(\xb_2,\xb_1,t) = \frac{1}{\sqrt{\det(-2\pi i\lm_3)} }
\exp \left[ -{i \over 2} \left( \xb_2\lm_3^{-1}\lm_4\xb_2
-2\xb_2\lm_3^{-1}\xb_1  +  \xb_1\lm_1\lm_3^{-1}\xb_1
\right) \right].
\label{prop}
\ee

The Hamiltonian can be re-expressed
in terms of creation and annihilation operators
via (\ref{rozhd})
\be
H \ = \ \frac {1}{2} \left(
\begin{array}{cc}
\ab & {\ab^{\dag}}
\end{array} \right)
\Bc (t) \left(
\begin{array}{c}
\ab \\ {\ab^{\dag}}
\end{array} \right),\qquad \Bc (t)=\ub^t\Bb (t)\ub .
\ee
Time-dependent Schr\"{o}dinger operators
 $\bb(t)$ and $\bb^{\dag}(t)$ of integrals of motion which coincide
at some zero time
with time-independent creation and annihilation operators
$\ab$ and $\ab^{\dag}$ were introduced in \cite{183.3}.
The former are expressed in terms of the latter via
the matrix $M$
\be
\left(
\begin{array}{c}
\bb \\ {\bb^{\dag}}
\end{array} \right) =
\left(  \begin{array}{cc}
\xi & \eta \\ \eta^* & \xi^*
\end{array} \right)
\left(
\begin{array}{c}
\ab \\ \ab^{\dag}
\end{array} \right)
\equiv M
\left(
\begin{array}{c}
\ab \\ \ab^{\dag}
\end{array} \right).
\ee
The matrix $M$ belongs to the group which is isomorphic to
the symplectic group.
Consequently the inverse transformation may be given by the formula
\be
\label{Ha_a_pro}
\left(
\begin{array}{c}
\ab \\ {\ab^{\dag}}
\end{array} \right) =
\left(  \begin{array}{cc}
\xi^* & -\eta \\ -\eta^* & \xi
\end{array} \right)
\left(
\begin{array}{c}
\bb \\ \bb^{\dag}
\end{array} \right) .
\label{trancrea}
\ee
Then the matrix $M$ can be found from the differential equation
\be
-i\dot M = M\Sigma\Bc
\ee
with the initial condition that $M$ is a unity matrix at zero time.
This equation follows from the condition that the operator $b(t)$ is
the integral of motion.
Below we will use the above expressions to calculate
the evolution of state vectors of parametric systems.

\section{Parametric cat state for one-mode oscillator}

The procedure described in Sec.~\ref{se:pasy}
is the easiest to apply to the case of one mode.
The Hamiltonian of the system under discussion has
the form (we use dimensionless variables)
\be
H = \frac{p^2}{2} + \Om^2(t) \frac{x^2}{2}.
\label{1ham}
\ee
In this case the transformation matrix (\ref{lambda})
is reduced to
\begin{eqnarray}
\lm_1 & = & \yp_r \equiv {\rm Re}\yp, \\
\lm_2 & = & -\dot\yp_r, \\
\lm_3 & = &  -\yp_i \equiv {\rm Im}\yp, \\
\lm_4 & = &  \dot\yp_i,
\end{eqnarray}
where $\yp$ is the solution of the classical oscillator equation
\be
\ddot\yp(t) + \Om^2(t)\yp(t) = 0
\ee
with the initial conditions ($\Omega(0) = 1$)
\be
\yp(0) = 1, \ \ \ \ \ \ \dot\yp(0) = i.
\ee
It is known (see for example \cite{183})
that the wave function of a coherent state
in coordinate representation is
\be
\label{cohpsy}
\langle x | \al \rangle = (\pi)^{-1/4} \exp \left[
-{x^2 \over 2} + \sqrt{2}\al x - {| \al |^2 \over 2 }
- {\al^2 \over 2} \right].
\ee
Acting by the propagator (\ref{prop}) on the
cat state (\ref{cat}) we obtain the wave function
to which the cat state evolves at time $t$
\be
\langle x | \alpha_+, t \rangle =
\langle x | 0, t \rangle
2N_+ \exp( - { |\al|^2 \over 2}
 - \frac {\alpha^2}{2}  {\yps \over \yp }  )
 \cosh
\left( { \sqrt{2} \alpha x \over \yp } \right),
\ee
\be
\langle x | \alpha_-, t \rangle =
\langle x | 0, t \rangle
2N_- \exp( - { |\al|^2 \over 2}
 - \frac {\alpha^2}{2}  {\yps \over \yp }  )
 \sinh
\left( { \sqrt{2} \alpha x \over \yp } \right),
\ee
where the evolution of the vacuum state is
\be
 \langle x | 0,t\rangle = (\pi)^{-1/4} \yp^{-1/2}
\exp \left( { i\dot\yp x ^2 \over 2\yp } \right)
\ee

The parameters
of transformations (\ref{trancrea})
 of annihilation and creation operators
in the one-mode case are
\be
\xi = \frac{ \yp - i\ypd}{2}, \ \ \ \ \ \ \ \ \ \ \
 \eta = \frac{-\yp - i\ypd} {2}
\ee
They allow us to obtain the mean values and dispersions for
the parametric oscillator.
The mean values of single creation or annihilation operator remain
equal to zero like for initial cat states.
The mean values of the pairs of operators are
\be
\langle \al_+| aa |\al_+ \rangle = \xi^{*2}\al^2 + \eta^2\al^{*2}
-\xi^*\eta -(\xi^*\eta+\eta^*\xi) |\al|^2\tanh |\al|^2
\ee
\be
\langle \al_+| a^{\dag} a |\al_+ \rangle
= -\eta^*\xi^{*}\al^2 + \xi\eta\al^{*2}
+|\eta|^2 +(|\xi|^2+|\eta|^2) |\al|^2\tanh |\al|^2
\ee
\be
\langle \al_-| aa |\al_- \rangle = \xi^{*2}\al^2 + \eta^2\al^{*2}
-\xi^*\eta -(\xi^*\eta+\eta^*\xi) |\al|^2\coth |\al|^2
\ee
\be
\langle \al_-| a^{\dag} a |\al_- \rangle
= -\eta^*\xi^{*}\al^2 + \xi\eta\al^{*2}
+|\eta|^2 +(|\xi|^2+|\eta|^2) |\al|^2\coth |\al|^2
\ee
These expressions allow to obtain the variances of the
coordinate and the momentum
\bear
\sigma(q,q) & = & \frac 12 + \sigma(a,a^+) +{\rm Re} \sigma(a,a) \\
\sigma(p,p) & = & \frac 12 + \sigma(a,a^+) -{\rm Re} \sigma(a,a) \\
\sigma(p,q) & = & {\rm Im} \sigma(a,a) .
\eear
Thus one can see that squeezing occurs only in one of the components,
e.g. in the momentum for real $\al$. For this case
\be
\sigma(p,p) = \frac 12 + |\eta|^{*2} + {\rm Re} (\xi^*\eta) +
|\al|^2 \tanh|\al|^2 |\xi + \eta|^2 - |\al|^2 {\rm Re} (\xi+\eta)^2
\ee
the variance will be bounded only for real $\xi$ and $\eta$,
and squeezing will be less than in the initial state.

\section{Parametric cat states for multimode oscillator}

The wave function of the multimode coherent state is
a straightforward generalization of (\ref{cohpsy}):
\be
\langle \xb | \Ab \rangle = \pi^{-N/4} \exp \left(
-{\xb^2 \over 2} + \sqrt{2}\xb\Ab - {|\Ab|^2 \over 2}
- {\Ab\Ab \over 2}
\right).
\ee
Acting by the propagator (\ref{prop})
on the multimode cat state (\ref{cat})
we obtain a wave function (compare with Eq.~(7.24) of Ref.~\cite{183.3})
for the squeezed cat states $|\alb_\pm,t\rangle$ which coincide
at the initial time with the states $|\Ab_\pm\rangle$
described in Section~1:
\begin{eqnarray}
\langle \xb | \alb_\pm, t \rangle =
\pi^{-N/4}\left[ \det  \lm_p\right]^{-1/2} \nonumber \\
\exp \left[
-\frac{1}{2} \xb \lm_p^{-1}\lm_q \xb -\frac{1}{2}
\Ab\lm_p^*\lm_p^{ - 1}\Ab - {|\Ab|^2 \over 2}
\right] \times  \nonumber \\ \times
2N_\pm
\begin{array}{c}
\cosh \\ \sinh
\end{array}
\left(\sqrt{2}\xb\lm_p^{-1}\Ab\right),
\end{eqnarray}
where $\lm_p = \lm_1-i\lm_3$, $\lm_q = \lm_4+i\lm_2$.

Further we obtain the photon distribution for the multimode
cat state. According to the first
approach we need to calculate
the overlap integral of the two
 Wigner
functions.
The first one corresponds to the  the multimode
squeezed coherent states
evolving due to parameter variation
$\rho = |\alb, t \rangle \langle \beb ,t|$.
The second one corresponds to the constant
coherent states $|\gmb \rangle \langle \deb |$.
The first Wigner function is expressed by the formula (8.5) of
Ref.~\cite{183.3}
\be
W_{\alb\beb} = 2^N \exp \left[ -\zb\Sigma_x\zb + 2\Ac\zb
 - \frac 12 \Ac\Sigma_x\Ac - \frac 12 |\Ac|^2 \right].
\ee
Here $\Sigma_x$ is the multidimensional analogue of one of the
Pauli matrices,
\be
\zb = \ub^{-1}\Qb(t), \ \ \ \ \ \ \Ac = \left( \begin{array}{c}
\beb^* \\ \alb
\end{array} \right),\ \ \ \ \ \
\Sigma_x = \left(
\begin{array}{cc}
\0b & \1b \\ \1b & \0b  \end{array} \right).
\ee
The second Wigner function is obtained from
the above Wigner function by changing from $\Qb(t)$ to $\qb$ and from
$\Ac$ to
\be
\Gamma = \left( \begin{array}{c}
\deb^* \\ \gmb
\end{array} \right).
\ee
The overlap integral gives the generating function for
the density matrix in Fock representation
\be
\langle \deb | \rho | \gmb \rangle = \exp \left( -{|\Gamma|^2 \over 2}
\right) \sum_{\mb,\nb=0}^\infty \frac { \deb^{*\mb}\gmb^\nb }
{ (\mb !\nb !)^{1/2} } \rho_{\mb\nb},
\ee
where $\nb = (n_1,n_2,... , n_N)$ is the vector
of photon numbers in all modes and $\nb ! = n_1 ! n_2 ! ... n_N !$.
{}From this decomposition we can find the photon number distribution
\be
P_\nb = \rho_{\nb\nb}.
\ee
The overlap integral is found to be
\be
\langle \deb | \rho | \gmb \rangle =
\exp \left[
- \frac{|\Gamma|^2}{2} - \frac 12 \Gamma R\Gamma
+ \Gamma R \yb
\right]
\tilde P(\Ac),
\ee
where
\be
R = \ub^{-1}(\Lambda^t\Lambda+1)^{-1}(\Lambda^t\Lambda-1)\ub
\Sigma_x
\ee
\be
\yb_i = -2\Sigma_x\ub^{-1}(\Lambda^t\Lambda-1)^{-1}\Lambda^t\ub^*\Ac_i
\ee
and the part bilinear in the coherent state vector $\Ac$ in the exponent is
\be
\tilde P(\Ac) =
{ 1 \over
 \sqrt{\det \left( \frac{1+\Lambda\Lambda^t}{2} \right) }
 }
\exp \left(
-\frac 12 |\Ac|^2 - \frac 12 \Ac\Sigma_x\Ac +
\Ac\ub^{-1}\Lambda(\Lambda^t\Lambda+1)^{-1}\Lambda^t\ub^*\Ac
\right).
\ee
Taking into account that the Gaussian distribution is
the generating function for Hermite polynomials
of $2N$ variables
and expressing the cat state through coherent states
(\ref{cat}), we obtain the photon distribution for
the cat state
\be
P_{\nb} = (1 \pm (-1)^{\sum_i n_i})^2 {N_\pm^2 \over \nb !} \left(
\tilde P(\Ac_1) \left[ H_{\nb\nb}^{\{ R\} } (\yb_1) \right]
\pm \tilde P(\Ac_2)  \left[ H_{\nb\nb}^{\{ R\} } (\yb_2) \right]
\right).
\ee
Here we choose
\be
\Ac_1 = \left(
\begin{array}{c}
\alb^* \\ \alb
\end{array} \right),
\ \ \ \ \ \ \ \
\Ac_2 = \left(
\begin{array}{c}
-\alb^* \\ \alb
\end{array} \right).
\ee
to be the parameter vectors corresponding to different signs of coherent
states in the cat state. Deriving the formula for the photon distribution
function we used the property of multimode Hermite polynomial to be either
even or odd function if the sum of indices of the polynomial is respectively
even or odd number. This property may be easily
proved from the definition of the Hermite polynomial trough the generating
function.

The second
approach to finding the photon distribution function
which gives equivalent but somewhat
simpler expressions is calculating the amplitudes of transitions
from initially squeezed \Scs \ to number states. For the squeezed
coherent
states these amplitudes are given in \cite{183.3} by the expression (5.58)
\be
\langle \mb | \alb, t \rangle _0 = \frac
{\exp \left( \frac 12 \alb \eta^*\xi^{-1}\alb
- {|\alb |^2 \over 2 }\right) }
{\mb !^{1/2} (\det\xi)^{1/2} } H_{\mb}^{\{\xi^{-1}\eta\} }
(\eta^{-1}\alb ).
\ee
Using the representation of a cat state in terms of coherent
states (\ref{cat}) we obtain the photon number distribution
\bear
(P_\mb)_0 = |\langle \mb | \alb_\pm ,t\rangle _0 |^2 \nonumber = \\
= (1 \pm (-1)^{\sum_i m_i} )^2 |N_\pm|^2 \frac
{\exp  \left( Re( \alb \eta^*\xi^{-1}\alb)
-|\alb|^2 \right) }
{\mb ! |\det\xi | }
| H_{\mb}^{\{\xi^{-1}\eta\} } (\eta^{-1}\alb )|^2.
\label{fla_osn}
\eear

Thus due to the constructive interference of the two amplitudes we have
the factor 4 and in the case of destructive interference we have the
factor zero in
the formula (\ref{fla_osn}). This formula demonstrates an example of the
wave interference in discrete configuration space of photon numbers.
It gives the nonzero probability to have even number of photons
only for even squeezed coherent state. In the multimode
even squeezed coherent state the probability
to have odd
number of photons is equal to zero. And vice versa this formula
gives the probability to have odd number of all the photons for the
odd coherent state.
The probability to have even number of photons in multimode
odd coherent state is equal to
zero. The same result follows from the above
formula for the photon distribution function expressed in terms of Hermite
polynomials with $2N$ indices.
By this method the photon number distribution is expressed through
$N$-index instead of $2N$-index Hermite polynomials.
It is easy to show that the found distribution function may be related to
the photon distribution in the multimode correlated state discussed in
Ref.~\cite{Ola}.
When the found probability is not equal to zero we have equality
\be
(P_\mb)_0 = 4|N_\pm|^2 \Pc_\mb
\ee
where the function $\Pc_\mb $ is the photon distribution
function for polymode squeezed and correlated light.
Due to this the envelope of the found distribution function if one
neglects the
fast oscillations connected with zeros at even or odd numbers
has the same shape that photon distribution function for polymode
squeezed and correlated light.
An example of such a single-mode evolution of originally
a cat state with $\al=2$ is represented in Fig.~1.
The frequency of the oscillator
(all further in dimensionless variables)
in (\ref{1ham}) is changing
according to
\be
\Om^2 = {1+ \kap \cos(2 t) \over 1 + \kap}
\ee
with $\kap=0.2$. The plot shows the envelope of the
photon distribution at different times.
{}From the plot and the above formulas one can see that
in the course of parametric excitation
the state conserves the cat properties
while the average photon number increases.

\section{Distribution function of \Scs\  subject to external forces}

Let us consider now the  evolution of \Scs\  distribution function due
to the influence of linear in qiuadratures terms in the Hamiltonian
of the multimode parametric oscillator.
\be
H=\frac {1}{2} \qb \Bb (t) \qb + \Cb \qb ,
\ee
where $2N$-vector $\Cb$ has c-number time-dependent components.
In terms of the occupation number operators the hamiltonian looks
similar to (\ref{Ha_a_pro})
\be
H \ = \ \frac {1}{2} \left(
\begin{array}{cc}
\ab & {\ab^{\dag}}
\end{array} \right)
\Bc (t) \left(
\begin{array}{c}
\ab \\ {\ab^{\dag}}
\end{array} \right)
+ \left(
\begin{array}{cc}
\ab & {\ab^{\dag}}
\end{array} \right)
\left(
\begin{array}{c}
\fb \\
\fb^*
\end{array}
\right).
\ee
Propagator for this system  (see, e.g., (5.58)
in Ref.\cite{183.3}) produces
the following transition  amplitued from a  coherent state
to a  number state (in the notation of Sec.~\ref{se:pasy} )
\be
\langle \mb | \alb, t \rangle =A_f \frac
{\exp \left( \frac 12 \alb \eta^*\xi^{-1}\alb
+\alb(\gmb^*-\eta^*\xi^{-1}\gmb)
-\frac 12 {|\alb|^2} \right) }
{\mb !^{1/2} (\det\xi)^{1/2} } H_{\mb}^{\{\xi^{-1}\eta\} }
(\eta^{-1}(\alb -\gamma )),
\label{coh_nu}
\ee
where $A_f$ is given by the formula
\be
A_f = \exp \left( \frac 12 \gmb \eta^* \xi^{-1} \gmb - {|\gmb|^2 \over 2}
 \right).
\ee
The $N$-vector $\gmb$ is the solution to the equation
\be
-i
\left(
\begin{array}{c}
\dot \gmb \\
\dot \gmb^*
\end{array}
\right)
= \Lambda \Sigma
\left(
\begin{array}{c}
\fb \\
\fb^*
\end{array}
\right),
 \ \ \ \ \ \gmb(0) = 0.
\ee
Transition amplitude from a coherent state $\alb$ to
a coherent state $\beb$ according to Eq.~(5.13) of Ref.~\cite{183.3} is
\bear
\langle \beb | \alb , t \rangle = (\det \xi)^{-1/2}
\exp [ - \frac 12 \beb \xi^{-1} \eta \beb
+ \beb\xi^{-1}(\alb - \gmb ) - \nonumber \\
- {|\al|^2 \over 2} + \frac 12 \alb \eta^* \xi^{-1}\alb + \alb
(\gmb^* - \eta^*\xi^{-1}\gmb ) + \frac 12 \gmb \eta^*\xi^{-1}\gmb
- {|\gmb|^2 \over 2} ]
\eear
is the generating function for the amplitudes (\ref{coh_nu}).

To obtain the quanta distribution function we
have to take the following sum
\be
P_\mb =
|N_{\pm}|^2
 |\langle \mb | \alb, t \rangle
\pm \langle \mb | -\alb, t \rangle |^2
\label{compdist}
\ee
In the previous case the amplitudes had the simple parity property, and due to
this we had the
 expression for the distribution function which had no cross
term. In the presence of linear terms in the interaction Hamiltonian
the amplitudes have no properties to be either even or odd functions of
$\alb $.
Due to this we will have four terms after factoring (\ref{compdist})
which are no longer different by merely a number factor.
Combining all four terms we have the final expression for the photon
distribution function in terms of multivariable Hermite polynomials.
The main difference from the previous distribution is that the
probability to have even and odd photon number is not equal to zero
now. So the property of even \Scs \ not to contain states with odd number
of photons  (and
vice versa) is destroyed. This property is the reason of high oscillations
of the photon distribution function for the \Scs . The squeezing in the
coherent component of \Scs \ produces the extra oscillations of the
photon distribution function. For squeezed and correlated states this
property of photon distribution function is known for one-mode case
\cite {Wheeler},\cite{Klimov} and for two-mode squeezed vacuum state
\cite{Caves},\cite {Man93}. It is obvious that in polymode squeezed even
and odd coherent states the photon distribution function preservs the
property to have strong oscillations. The linear terms in the interaction
Hamiltonian reduce the oscillations related to the interference of the
two squeezed components of the \Scs \ producing nonzero probability
to have both even and odd number of photons.

For small
values of this parameter we can expand
the transition amplitude and
find the first nonzero correction to the distribution
obtained in the previuos section. For this we differentiate
the generating function, take it at $\gmb = 0$, and multiply
by $\gmb$. The coefficients of
expansion of this expression into  a power
series over $\gmb$ are the corrections to the transition amplitudes
\be
\Delta \langle \mb | \alb , t \rangle =
(\alb \gmb^* - \alb \eta^*\xi^{-1}\gmb ) \langle
\mb | \alb ,t \rangle _0 -
\sum_{i,j=0}^N \langle \mb /i| \alb , t \rangle _0
\sqrt{m_i} (\xi^{-1})_{ij}\gmb_j,
\ee
where a photon number vector
$ \mb /i = (m_1, \dots , m_i - 1, \dots , m_N)$
exists when $m_i > 0$. The corrections to the
squeezed \Scs \ amplitudes and the photon number
distribution are
\be
\Delta \langle \mb | \alb_\pm , t\rangle =
N_\pm
\Delta \langle \mb | \alb , t \rangle
(1 \mp (-1)^{\sum_i m_i} )
\ee
\be
\Delta P_{\mb} = \langle \alb_\pm, t| \mb \rangle _0
\Delta \langle \mb | \alb_\pm , t \rangle + c.c.
\ee
They contribute at the photon numbers other than
those comprised by the original even or odd state.
So we demonstrated that decoherence of \Scs \ due to linear
terms in the interaction Hamiltonian tends to remove strong
oscillations of distribution function existing in the
\Scs \ initially
squeezed by pure quadratic Hamiltonian.
These terms add the contribution
of opposite parity.

The evolution of a free particle when the
returning force in (\ref{1ham}) is
\be
\Om (t) = 0
\ee
under the action of
an external force $\fb = 1$ is represented in Fig.~2.
In the plot
one can see how the high oscillations become smoother
when the parameter $\gmb (\fb )$ is different from zero.
The evolution of the bounded particle
\be
\Om (t)= 1
\ee
under the action of the force $\fb = 0.2$
is shown in Fig.~3.
In the course of evolution the odd photon numbers
instead of the even ones begin to prevail.
On the other hand the state is not transferred to a really
odd cat state, but the photon distribution becomes
less oscillating.

\section{Conclusion}
We have shown that the photon distribution function for
the multimode squeezed \Scs \ differs essentially from the product
of Poisson distribution functions describing the multimode coherent
state. In the limiting  case of the symplectic transform described by the
unity element, the above expressions for the photon distribution functions
coincides with the distribution function found in Ref.~\cite{Ansari}.
There exists interaction which destroys the propery of \Scs \ to have
only even or only odd numbers of photons. The stronger the influence of this
interaction, the more contribution of the states with opposite pairity
of photons number is.

\newpage

\begin{figure}
\caption{ The envelope of the photon distribution
of the even or the odd cat state with $\alb = 2$
evolving under the parameter variation (69).
The curves in the order of the lowering maxima correspond
respectively to the times
$t=0., \ 2., \ 4.5, \ 6.$}
\end{figure}

\begin{figure}
\caption{The evolution of the initially even cat state
under the action of the external driving force
in the absence of the returning force (80).
The photon number distributions are shown for
the times $t = 0., \ 0.2, \ 0.4, \ 0.6$ in the
order of less value at $n=4$. }
\end{figure}

\begin{figure}
\caption{The evolution of the initially even cat state
under the action of the external driving force
with constant oscillator frequency (81).
The photon number distribution is shown for
a) the times $t=0., \ \pi/4, \ \pi/2$
in the order of lowering maxima;
b) the times
$t= \pi/2, \ 3\pi/4, \ \pi$
in the order of higher value $P(0)$.}
\end{figure}

\end{document}